\documentclass[twocolumn,prl]{revtex4}

\usepackage{graphicx}
\usepackage{dcolumn}
\usepackage{bm}
\usepackage{multirow}
\usepackage{color}
\usepackage[normalem]{ulem}
\usepackage{amsmath}
\usepackage{gensymb}
\usepackage[colorlinks=true]{hyperref}

\begin{document}

\title{Controlling interlayer excitons in MoS$_2$ layers grown by chemical vapor deposition}

\author{Ioannis Paradisanos$^{1*}$}
\author{Shivangi Shree$^{1*}$}
\author{Antony George$^2$}
\author{Nadine Leisgang$^3$}
\author{Cedric Robert$^1$}
\author{Kenji Watanabe$^4$}
\author{Takashi Taniguchi$^4$}
\author{Richard J. Warburton$^3$}
\author{Andrey Turchanin$^{2,5}$}
\author{Xavier Marie$^1$}
\author{Iann C. Gerber$^1$}
\email{igerber@insa-toulouse.fr}
\author{Bernhard Urbaszek$^1$}
\email{urbaszek@insa-toulouse.fr}

\affiliation{\small$^1$Universit\'e de Toulouse, INSA-CNRS-UPS, LPCNO, 135 Avenue Rangueil, 31077 Toulouse, France}
\affiliation{\small$^2$Friedrich Schiller University Jena, Institute of Physical Chemistry, 07743 Jena, Germany}
\affiliation{\small$^3$Department of Physics, University of Basel, Basel, Switzerland}
\affiliation{\small$^4$National Institute for Materials Science, Tsukuba, Ibaraki 305-0044, Japan}
\affiliation{\small$^5$Abbe Centre of Photonics, 07745 Jena, Germany}

\begin{abstract} Combining MoS$_2$ monolayers to form multilayers allows to access new functionalities. In this work, we examine the correlation between the stacking order and the interlayer coupling of valence states in MoS$_2$ homobilayer samples grown by chemical vapor deposition (CVD) and artificially stacked bilayers from CVD monolayers. We show that hole delocalization over the bilayer is allowed in 2H stacking and results in strong interlayer exciton absorption and also in a larger A-B exciton separation as compared to 3R bilayers, where both holes and electrons are confined to the individual layers. Comparing 2H and 3R reflectivity spectra allows to extract an interlayer coupling energy of about $t_\perp=49$~meV. Obtaining very similar results for as-grown and artificially stacked bilayers is promising for assembling large area van der Waals structures with CVD material, using interlayer exciton absorption and A-B exciton separation as indicators for interlayer coupling. Beyond DFT calculations including excitonic effects confirm signatures of efficient interlayer coupling for 2H stacking in agreement with our experiments.
\end{abstract}


\maketitle

\begin{figure*}
\includegraphics[width=0.85\linewidth]{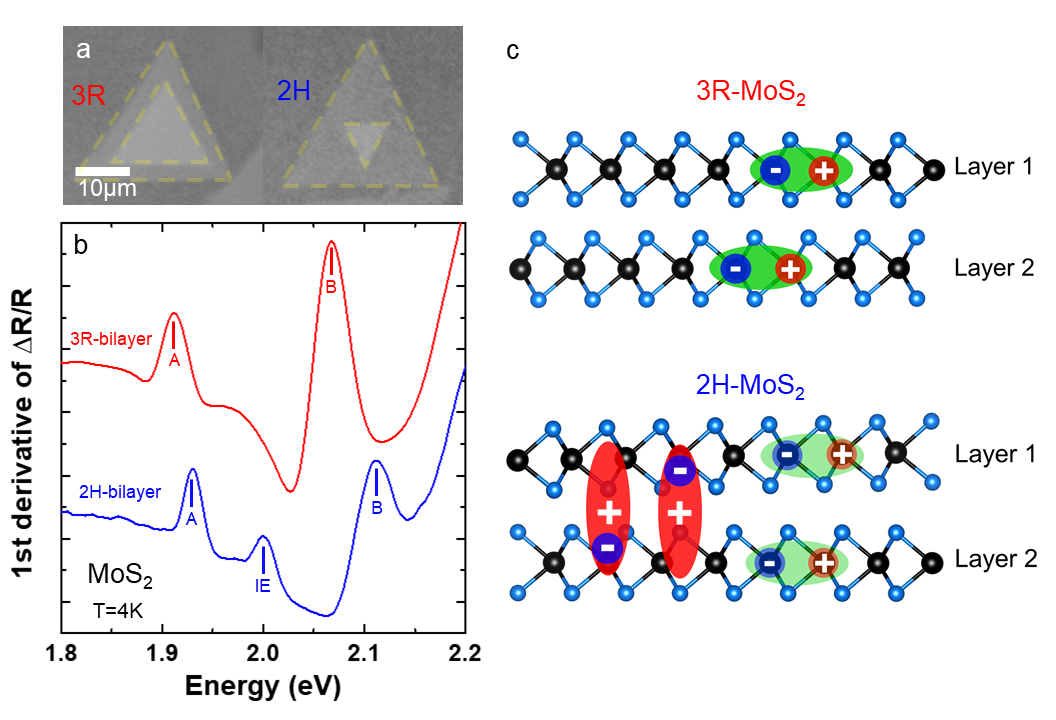}
\caption{\label{fig:fig1} \textbf{Spectroscopy of CVD grown bilayers encapsulated in hBN.} (a) Optical microscope images of as-grown 3R (left) and  2H CVD MoS$_2$ bilayers (right) on SiO$_2$/Si before pick-up.  (b) First derivative of white light reflection spectrum  for as-grown 2H-bilayer (blue) and as-grown 3R-bilayer (red), recorded at T$=4$~K, both bilayers are encapsulated in high quality hBN for optical spectroscopy \cite{Taniguchi:2007a}. (c) Schematic of 3R stacked bilayer with intralayer excitons (top) compared to 2H stacked bilayer where in addition interlayer excitons are observed as in panel (b).}
\end{figure*}

\textbf{Introduction.---} Transition metal dichalcogenides (TMDs) with the form MX$_2$ (M = Mo, W, Ti, etc and X = S, Se, Te) have tunable electronic properties from metallic to semiconducting depending on the crystal symmetry, composition and number of layers \cite{Novoselov:2016a,Mak:2016a,Schaibley:2016a,unuchek2018room,Schneider2018a,Koperski:2017a,dufferwiel2017valley,Scuri:2018a,hong2014ultrafast}. The band structure of TMD semiconductors is drastically modified by changing the sample thickness by just one atomic monolayer \cite{Splendiani:2010a,Mak:2010a,tonndorf2013photoluminescence}. For instance, the combination of two different monolayer materials such as MoSe$_2$-WSe$_2$ into a heterobilayer results in type II band alignment and opens new research perspectives on periodic moir\'e potentials for carriers in the different layers and the resulting interlayer excitons \cite{doi:10.1021/acs.nanolett.8b03266,tran2019evidence,jin2019observation,seyler2019signatures,alexeev2019resonantly}. Twisted homobilayers of graphene, WSe$_2$ and MoSe$_2$ allow accessing new superconducting phases and correlated insulating states \cite{cao2018unconventional,wang2019magic,shimazaki2019moir}.\\
\indent  To access the new functionalities provided by assembling monolayers to form multilayers it is necessary to identify physical parameters that strongly depend on interlayer coupling and to experimentally control them. Our approach is to compare CVD grown MoS$_2$ bilayers with artificially stacked bilayers made from CVD monolayers. We study for both as-grown CVD and individually assembled cases the 2H (180$^{\circ}$ twist angle, see schematic in Fig.\ref{fig:fig1}c) and 3R (0$^{\circ}$ twist angle) stacking which gives precise control over the interlayer coupling and hence interlayer exciton formation. Studying these two precise alignments is also relevant for samples initially assembled with other twist angles as reconstruction results also in these experiments in the formation of  $\mu$m  wide  2H and 3R areas \cite{Weston:2019reconstr,sung2020broken}, which are energetically most stable. We show that the valence states for 2H bilayers are strongly impacted by interlayer coupling as the hole is delocalized over the 2 layers \cite{Slobodeniuk:2019fine,gerber:2019interlayer,deilmann2018interlayer}. This results in important changes in the optical spectra governed by K-K transitions as we observe strong absorption from interlayer excitons and a clear change in separation between A- to B-exciton transition in differential white light reflection at T$=4$~K. These observations are made possible due to the drastically improved optical quality of CVD samples removed from the growth substrate and encapsulated in hBN \cite{Shree:2019hom}. We show that both indicators for interlayer coupling are absent in the measured 3R bilayer spectra as hole hopping between the layers is symmetry forbidden \cite{gong2013magnetoelectric}. Comparing for 3R (no interlayer coupling) and for 2H the A-B exciton absorption spectra allows us to extract an experimental value of the perpendicular hopping (coupling) term of $t_\perp\approx49~$meV, important for moir\'e superlattices \cite{Tong:2016kh} and so far only roughly estimated from theory \cite{gong2013magnetoelectric}. To artificially stack large area CVD layers and control interlayer coupling through stacking ({\textit{i.e.}} 0$^\circ$ or 180$^\circ$ twist angle) is technologically relevant for 2D materials optoelectronics \cite{alexeev2019resonantly}, as CVD substrates are covered by a large number of monolayers and are very practical to stack (twist) due to their symmetric triangular shape and well characterized edge termination.\\
\indent In addition to our optical spectroscopy experiments we show in density functional theory (DFT-$GW$) calculations that at the independent particle and the quasiparticle (exciton) level description of the system the valence band splittings for 2H as compared to 3R are different due to interlayer coupling.  In our calculated absorption spectra with excitonic effects solving the Bethe-Salpeter-Equation (BSE) for 2H stacking we show strong interlayer exciton absorption, absent for 3R stacking.

\textbf{Interlayer excitons in as-grown CVD MoS$_2$ homobilayers.---} The thermodynamically most stable configurations of TMD homobilayers are the 2H and the 3R stacking \cite{Grimme:2010ij,gerber:2019interlayer}. In practice, most naturally occurring molybdenite shows 2H not 3R stacking. In this work we focus on high quality CVD grown flakes for several reasons : during CVD growth of MoS$_2$ both 2H and 3R stackings for bilayers can occur \cite{xia2015spectroscopic} and we are therefore able to compare the optical response for samples grown under identical conditions. Secondly, as all CVD flakes on our substrate show the same edge termination, we can artificially stack two layers in 2H and 3R configuration to compare with the as grown samples - see discussion below. Third, many monolayers cover the SiO$_2$ substrate and can all be picked-up in a single step, which makes fabrication of bilayer structures very efficient.\\
\begin{figure*}
\includegraphics[width=0.9\linewidth]{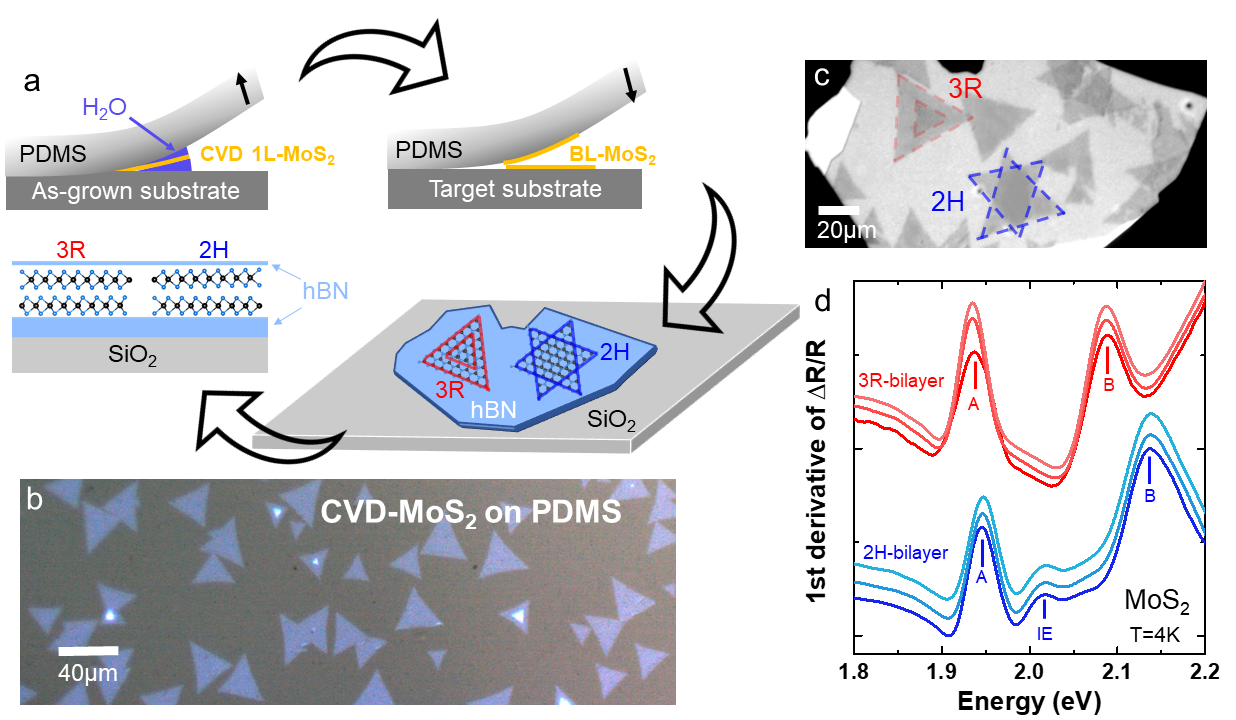}
\caption{\label{fig:fig2} \textbf{Artificial stacking of CVD monolayers into bilayers.} (a) Schematic of sample pick-up, bilayer assembly and encapsulation for optics. (b) Optical micrograph of CVD-grown MoS$_2$ monolayers and a few homobilayers, transferred to the PDMS stamp following water-assisted pick-up \cite{Jia:2016water} from the growth substrate. (c) Artificially-assembled 3R and 2H MoS$_2$ homobilayers, fabricated by an all-dry deterministic transfer process. (d) First derivative of reflectivity spectra collected from three different areas of the artificially stacked 3R (red) and 2H (blue) MoS$_2$ homobilayers, shown in (c). Spectra have been shifted for clarity.}
\end{figure*}
\indent Optical microscope images of \textit{as-grown} CVD bilayers on SiO$_2$/Si with 3R and 2H stacking are presented in Fig.\ref{fig:fig1}a. The stacking can be determined already by the relative rotation of the triangular monolayers and was confirmed in second harmonic generation (SHG) experiments. The SHG signal for 2H stacking was not detectable (inversion symmetry restored) but we could perform detailed angle dependent SHG for the 3R stacking (broken inversion symmetry), see supplement. The high quality MoS$_2$ bilayers and monolayers were grown by a modified CVD process in which a Knudsen-type effusion cell is used for the delivery of sulfur precursor \cite{George_2019}.  Using a water-assisted pick-up technique \cite{Jia:2016water}, the as-grown CVD bilayers have been deterministically transferred and encapsulated in hBN to achieve high optical quality \cite{Shree:2019hom}, which has recently been shown to be crucial for optical spectroscopy on CVD samples lowering the typical emission linewidth from about 50~meV to below 5~meV at T$=4$~K. The thickness of the top and bottom hBN has been carefully selected to optimize the oscillator strength of the interlayer excitons (IEs) \cite{gerber:2019interlayer}. After encapsulation, the samples were cooled down to T$=4~$K in a closed-cycle cryostat and a series of differential reflectivity measurements with a home-built confocal microscope have been performed at different locations of the samples, see methods. We define differential reflectivity as $(R_{sam}-R_{sub})/R_{sub}$, where $R_{sam}$ is the intensity reflection coefficient of the sample with the MoS$_2$ layers and $R_{sub}$ is the same structure without the MoS$_2$. Note that the overall shape and amplitude of the differential reflectivity signal also depends on cavity effects (thin-layer interference) given by top and bottom hBN and SiO$_2$ layer thickness (see \cite{Robert:2018a} for details).
In Fig.\ref{fig:fig1}b, the first derivative of the differential reflectivity spectra for as-grown CVD 2H and 3R MoS$_2$ bilayers can be compared.
There are two striking differences between the 2H and 3R bilayer spectra: \textbf{(i)} While A and B intralayer excitons are identified for both configurations, a pronounced feature at $\approx$ 2~eV appears exclusively in the 2H bilayer. This feature is assigned to an interlayer state, its energy being in good agreement with the very recently identified IEs in high quality and hBN encapsulated \textit{exfoliated} MoS$_2$ bilayers with 2H stacking \cite{gerber:2019interlayer,Slobodeniuk:2019fine,niehues2019interlayer}, in contrast to CVD grown samples studied here. This observation of IEs was made possible by our specific CVD sample preparation for the optical spectroscopy experiment \cite{Shree:2019hom}. IE absorption was not detectable in very detailed earlier works due to considerably larger optical linewidth or detection of emission and not absorption \cite{shinde:2018stacking,huang2014probing,yeh2016direct,van2014tailoring}. In contrast to 2H stacking, in the 3R configuration no additional states are detected between the A - and B-excitons, thus indicating that delocalization of holes is not allowed in this particular stacking order \cite{gerber:2019interlayer}, see below for a more detailed discussion. \textbf{(ii)} The separation between the A -  and B -  exciton transitions is considerably larger in the 2H bilayers (about 185 meV) as compared to the 3R bilayer (about 150 meV, mainly given by the spin-orbit splitting in the valence band). This is a second indication for efficient interlayer coupling of A-B valence states for 2H stacking, as the separation of the valence states mainly governs the A-B exciton separation \cite{gong2013magnetoelectric,Slobodeniuk:2019fine,Kormanyos:2015a}.

\textbf{Control of the interlayer coupling in artificially-assembled MoS$_2$ homobilayers.---}
In Fig.\ref{fig:fig1} we show that as-grown CVD MoS$_2$ bilayers experience interlayer coupling resulting in interlayer exciton formation, here observed for a non-contaminated interface between the top and bottom layer. Contamination from secondary transfer processes could potentially suppress the coupling between the layers and hence IE formation. By choosing 2H or 3R orientation manually when stacking CVD monolayers to form bilayers one can allow hole tunneling between the layers or not. This  requires to pick-up the CVD monolayers from their growth substrate while maintaining their structural integrity, optical quality and a sufficiently clean interface after transfer. Furthermore, fine control of the twist angle between the top and bottom layer is needed, since the IE formation is allowed only in a precise stacking order, see sample preparation schematic in Fig.\ref{fig:fig2}a. Here we use water-assisted deterministic transfer that allows the ability to controllably assemble CVD bilayers  with a desired twist angle \cite{Jia:2016water}. First, CVD monolayers have been carefully picked-up from the growth substrate and transferred to PDMS (Fig.\ref{fig:fig2}b) \cite{Jia:2016water,Gomez:2014a}. The structural integrity of the CVD monolayers is preserved in this case and the following step is to slowly assemble 2H and 3R bilayers and encapsulate them in hBN as shown in Fig.\ref{fig:fig2}c. Small deviations from $0^{\circ}$ or $180^{\circ}$ twist angle are expected but natural reconstruction of the bilayer will again favor the lowest energy arrangement, 3R and 2H, respectively \cite{Weston:2019reconstr,sung2020broken}.\\
\indent Differential reflectivity spectra have been collected from ten different areas of the assembled 2H and 3R bilayers of Fig.\ref{fig:fig2}c. In Fig.\ref{fig:fig2}d, three typical examples of the assembled 2H and 3R spectra are presented. The spectra show a striking resemblance with the as-grown bilayer spectra discussed before in Fig.\ref{fig:fig1}b. So also for the assembled 2H bilayers we identify  clear interlayer exciton absorption and an increased separation between the A- and B-excitons. It is important to note that the IE transition was clearly observed over the whole surface area of the manually constructed 2H bilayer. We take it as a strong indication of efficient interlayer coupling and possibly efficient reconstruction/self-rotation to the 2H configuration. We therefore further confirm the formation of IEs exclusively in the 2H stacking. By manually choosing the stacking configuration i.e. twist angle, it is possible to tune the valence band splitting and the formation of interlayer excitons in large area, high quality CVD samples. \\
\begin{figure*}
\includegraphics[width=0.95\linewidth]{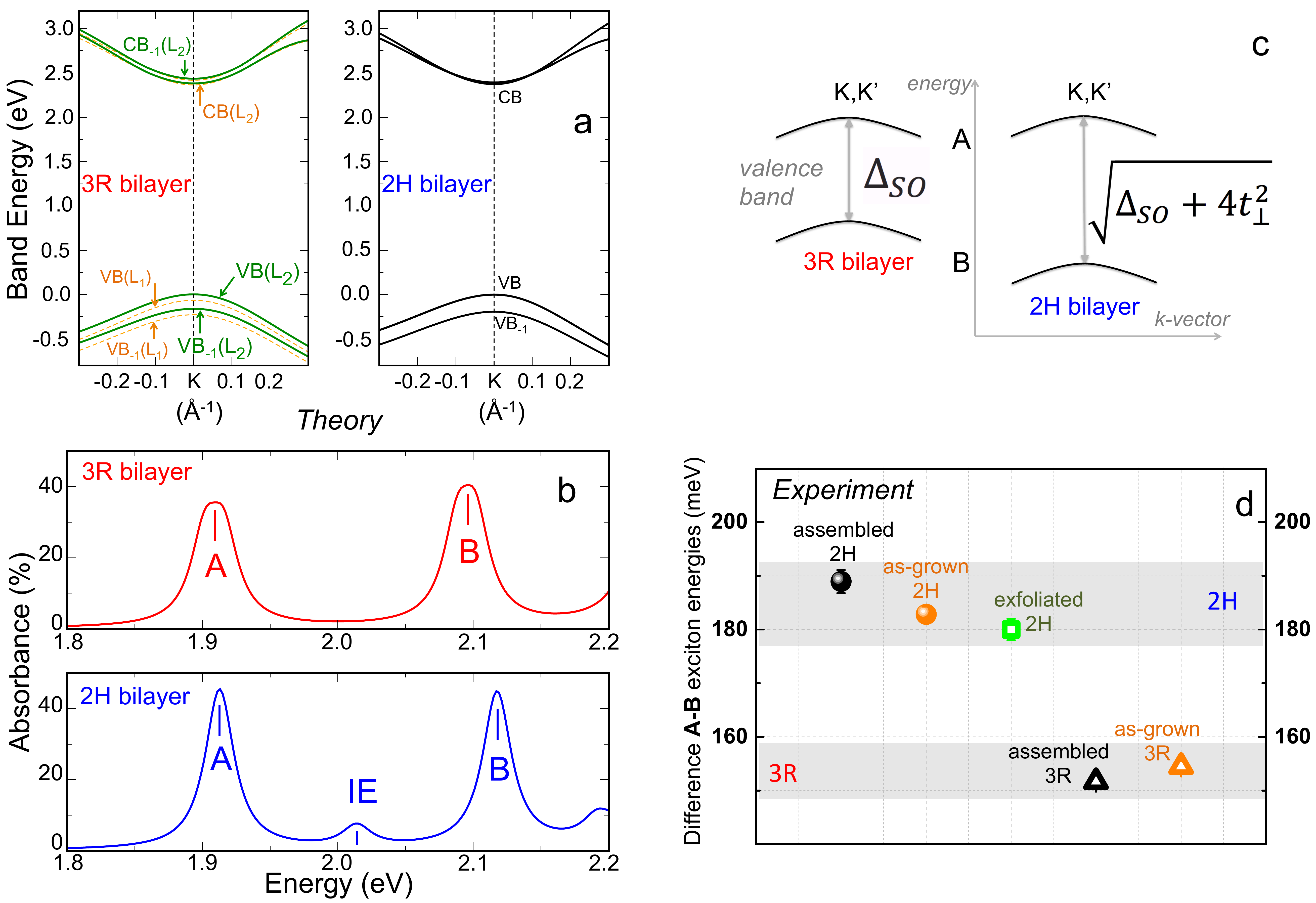}
\caption{\label{fig:fig3} \textbf{Interlayer coupling in theory and experiment.}  (a) Valence and conduction bands around K-point calculated with DFT-GW for 2H and 3R stacking, with the energy value of the highest VB set to 0. (b) Calculated absorption using DFT-GW-BSE for both stackings. (c) Schematic of the A- and B valence bands for 3R bilayers (left) and 2H  bilayers (right) as a function of the Spin-Orbit splitting $\Delta_{SO}$ and the interlayer coupling parameter $t_{\perp}$. (d) Energy difference between $B$ and $A$ exciton for the as-grown (orange), as well as artificially-assembled (black) 2H and 3R MoS$_2$ homobilayers. The error bars represent the standard deviation extracted over 10 different spectra in each case. }
\end{figure*}

\textbf{Beyond-DFT band structure calculations and exciton absorption.---} In addition to optical spectroscopy we perform beyond DFT calculations to study the striking differences between 2H and 3R MoS$_2$ bilayers, see methods for the computational details. 
Please note that our $GW$+BSE calculations are performed for MoS$_2$ bilayers in vacuum for simplicity and not in hBN. The general target of our calculations is to understand the microscopic origin of the optical transitions and to reproduce the energetic order qualitatively. In $GW$ calculations we compare band structures corrected by screening effects, the exciton description being added later. In the vicinity of the K-point of the Brillouin zone, 
see Fig.\ref{fig:fig3}a and schematic in Fig.~\ref{fig:fig3}c, differences between 2H and 3R stacking in VB, VB$_{-1}$ states are clear. We find a VB splitting in the 2H bilayer that is 19~meV larger than in the 3R bilayer. By solving the BSE we obtain the absorption for the 2H and 3R bilayers shown in Fig.~\ref{fig:fig3}b. The main characteristics are (i) the presence of a strong interlayer exciton peak in 2H and (ii) a larger A-B exciton separation for 2H than for 3R configuration, exactly as found in the experiments in Figs.\ref{fig:fig1} and \ref{fig:fig2}. Interestingly, in 3R stacking, there is a minor departure from degeneracy that splits VB and VB$_{-1}$  states from distinct layers when they remain degenerate in 2H configuration. As a consequence, two distinct A-type excitons separated by only 14~meV constitute the A-peak of the 3R-bilayer absorption spectrum in our calculation, explaining its larger width compared to the 2H case, see Fig.\ref{fig:fig3}c.

\textbf{Discussion.---}\\
To summarize the main experimental findings, \textbf{first} we observe strong interlayer exciton absorption between the main A - and B - exciton transitions for CVD grown (Fig.\ref{fig:fig1}b) and artificially stacked (Fig.\ref{fig:fig2}d) 2H bilayers, whereas this interlayer transition is absent for 3R stacking as hole tunnelling is symmetry forbidden and only intralayer exciton transitions are observed. A \textbf{second} striking observation is that stacking of the layers also affects the energy difference between A - and B - exciton transitions. This is demonstrated in Fig.\ref{fig:fig3}d, where the A - B exciton energy difference is compared between as-grown and assembled 2H and 3R bilayers. It is apparent that 2H bilayers exhibit a significantly larger energy difference between the A - and B - exciton states, compared to 3R bilayers. \\
\indent Our next target is to experimentally extract the interlayer hopping term, $t_\perp$ based on a $k . p$ model of bilayers in the vicinity of $K$ points \cite{gong2013magnetoelectric,Tong:2016kh,Slobodeniuk:2019fine} and compared it to post-DFT estimates.  As indicated in Fig.~\ref{fig:fig3}c, for 3R stacking the measured A- to B- exciton splitting $S_{\text{3R}}$ is roughly given by the spin orbit splitting as $S_{\text{3R}} = \Delta_\text{SO}$. For 2H-stacking the A-B splitting in the valence band depends in addition on the coupling energy $t_\perp$ as  $S_{\text{2H}} = \sqrt{\Delta_\text{SO}^2+4t_\perp^2}$ and hence
\begin{equation}\label{eq:coupling}
t_\perp = \sqrt{\frac{S_{\text{2H}}^2-\Delta_\text{SO}^2}{4}},
\end{equation}
where $S_{\text{2H}}$ is the measured A  - B exciton splitting of the as-grown 2H MoS$_2$ bilayer and as a value for $\Delta_\text{SO}$ we take the measured A-B separation in the 3R sample. For $S_{\text{2H}}=183$~meV and $\Delta_\text{SO}=155$~meV, we obtain $t_\perp \approx 49$~meV. This value can be compared to the ones extracted from our standard DFT as previously done~\cite{gong2013magnetoelectric}, or from more advanced $GW$ and $GW$+BSE calculations that correct for screening effects and include excitonic effects. Table~\ref{tab:tab-split} summarizes calculated valence band splittings and A-B energy differences for monolayer, 2H and 3R stacking as well as the corresponding coupling strength, directly extracted from the VB splitting
as in \cite{Tong:2016kh} or Eq.~\ref{eq:coupling}. The agreement between theoretical, including previous rough estimates~\cite{gong2013magnetoelectric}, and experimental results is good as we reproduce the larger A-B splitting for the 2H bilayers as compared to 3R.
Although we measure exciton transitions and not directly the valence band splitting in the 2H bilayer, the agreement between theory and experiment strongly supports our interpretation for the reason behind the different A-B exciton splitting. It should be noted that for $3R$ bilayers, $t_\perp = 0$ since interlayer hopping is not allowed in this case. This short numerical analysis highlights that the efficiency of this interlayer coupling will depend on the ratio of $t_\perp$ versus the spin-orbit valence band splitting \cite{PhysRevB.97.241404}, which is much smaller in MoS$_2$ ($\Delta_{SO}\approx 150~$meV) as compared to WSe$_2$ ($\Delta_{SO}\approx 430~$meV). This leads in principle to tunable interlayer coupling for MoS$_2$ \cite{wu2013electrical} and so called spin-layer-locking for WSe$_2$ bilayers \cite{Jones:2014a}.\\
\indent The A-B exciton splitting in MoS$_2$ bilayers has been previously studied by several groups  \cite{shinde:2018stacking,latzke:2015electronic,du2018:temperatureic,zhang:2015valence,jin:2013direct}. In these reports, the spin-orbit coupling and interlayer coupling have been discussed, but neither interlayer exciton formation nor an experimental analysis of the coupling term $t_\perp$. In our study the direct comparison in the same set-up on the same substrate of 2H and 3R bilayers with good optical quality due to encapsulation allows to determine the difference in A - B exciton energies precisely, which we ascribe to interlayer coupling, as supported by our quasi-particle $GW$ and exciton absorption calculations. From an experimental point of view, our results suggest two practical test criteria for interlayer coupling following artificial stacking : the strong interlayer exciton absorption and the clear difference in A-B exciton transition energies.  The physics discussed here for 2H and 3R bilayers is also relevant for samples with a twist angle slightly different from 0$^{\circ}$ or 180$^{\circ}$ as reconstruction/self-rotation results in artificial stacks typically in large areas of 2H and 3R stacking, which will show the optical properties of the samples investigated here. 

\begin{table*}
\begin{center}
\begin{tabular}{lcccccccc}
\hline
\hline
 & & ML & & 3R-bilayer & & 2H-bilayer & & $t_\perp$\\
\hline 
VB splitting & & 178 (189) & &175 (189) & & 194 (203) & & 57 (42) \\
$S$ & & 185 & & 186 & & 205 & & 43\\
\hline
\hline
\end{tabular}
\end{center}
\caption{\label{tab:tab-split} Valence band splittings, A-B transition energy differences ($S$) extracted from $GW$ and $GW$+BSE calculations and the corresponding interlayer coupling parameters. Values extracted for standard DFT calculations are in parentheses. 
All values are given in meV. }
\end{table*}

\textbf{Methods}.\\
 \textbf{CVD samples growth.-} MoS$_2$ crystals were grown on thermally oxidized silicon substrates (Siltronix, oxide thickness 300~nm, roughness $< 0.2$~nm RMS) by a modified CVD growth method as described in Ref.~\cite{George_2019} in detail.\\
 \textbf{Sample pick-up and encapsulation.-} A clean PDMS stamp was first placed on a glass slide and the SiO$_2$/Si substrate containing the as-grown CVD MoS$_2$ monolayers and homobilayers was brought in contact with the PDMS stamp \cite{Jia:2016water}. The substrate was pressed against the PDMS stamp and distilled water droplets were injected at the perimeter of the substrate. Water droplets penetrated into the SiO$_2$/MoS$_2$/PDMS interface and after 1 minute the SiO$_2$/Si substrate was carefully lifted, resulting into the transfer of a large area of CVD-grown MoS$_2$ triangles onto the PDMS stamp, as shown in Fig.~\ref{fig:fig2}b. Then the PDMS stamp (with the CVD-grown MoS$_2$ on top) was dried off using a nitrogen gun. Finally, hBN flakes were exfoliated from high quality bulk crystal \cite{Taniguchi:2007a} onto the target substrate and subsequent deterministic-dry transfer of the CVD-grown MoS$_2$ triangles from the PDMS stamp on top of the hBN was applied.\\ 
 \textbf{Optical spectroscopy Set-up.-}  Low temperature reflectance measurements were performed in a home-built micro-spectroscopy set-up assembled around a closed-cycle, low vibration attoDry cryostat with a temperature controller ($T=4$~K to 300~K).  The white light source for reflectivity is a halogen lamp with a stabilized power supply focussed initially on a pin-hole that is imaged on the sample. The emitted and/or reflected light was dispersed in a spectrometer and detected by a Si-CCD camera. The excitation/detection spot diameter is $\approx1\mu$m, i.e. smaller than the typical size of the homobilayers. \\ 
\textbf{Methods for DFT and $\boldsymbol{GW}$ calculations.-} \\  
The atomic structures, the quasi-particle band structures and optical spectra have been obtained from DFT calculations using the VASP 
package \cite{Kresse:1993a,Kresse:1996a}. The plane-augmented wave scheme \cite{blochl:prb:94,kresse:prb:99}  has been used to to treat core electrons. We have set the lattice parameter value of 3.22~\AA~ for all the runs. A grid of 15$\times$15$\times$1 k-points has been used, in conjunction with a vacuum height of  21.9~\AA, for all the calculation cells. The geometry's optimization process has been performed at the PBE-D3 level~\cite{Grimme:2010ij} in order to include van der Waals interaction between layers. All the atoms were allowed to relax with a force convergence criterion below $0.005$ eV/\AA. Heyd-Scuseria-Ernzerhof (HSE) hybrid functional~\cite{heyd:jcp:04_a,heyd:jcp:05,paier:jcp:06} has been used as approximation of the exchange-correlation electronic term, including SOC, to determine eigenvalues and wave functions as input for the full-frequency-dependent $GW$ calculations~\cite{Shishkin:2006a} performed at the $G_0W_0$ level.  An energy cutoff of 400 eV and a gaussian smearing of 0.05 eV width have been chosen for partial occupancies, when a tight electronic minimization tolerance of $10^{-8}$ eV was set to determine with a good precision the corresponding derivative of the orbitals with respect to $k$ needed in quasi-particle band structure calculations. 
The total number of states included in the $GW$ procedure is set to 1280, in conjunction with an energy cutoff of 100 eV for the response function, after a careful check of the direct band gap convergence (smaller than 0.1 eV as a function of k-points sampling). Band structures have been obtained after a Wannier interpolation procedure performed by the WANNIER90 program~\cite{Mostofi:2008ff}.
All optical excitonic transitions have been calculated by solving the Bethe-Salpeter Equation~\cite{Hanke:1979to,Rohlfing:1998vb}, using the twelve highest valence bands and the sixteen lowest conduction bands to obtain eigenvalues and oscillator strengths on all systems. From these calculations, we report the absorbance values by using the imaginary part of the complex dielectric function.

\section{Supplementary Information}

\subsection*{Second Harmonic Generation}

\begin{figure}
\centering
\includegraphics[width=80mm]{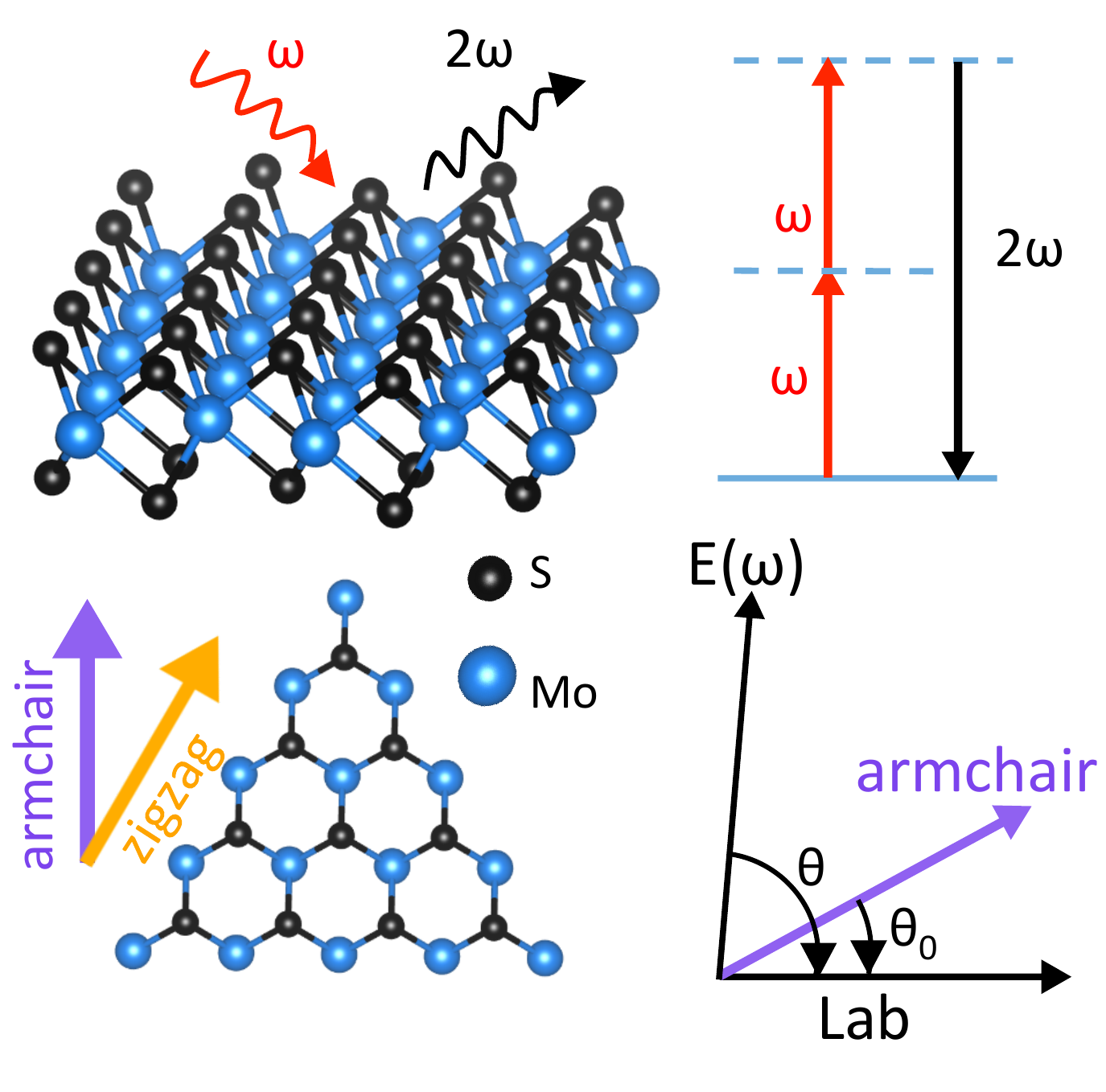}\caption{Schematic representation of the crystallographic orientation of MoS$_2$ monolayer and the polarization of the SHG.}
\label{FigS1}
\end{figure}

\begin{figure}
\centering
\includegraphics[width=90mm]{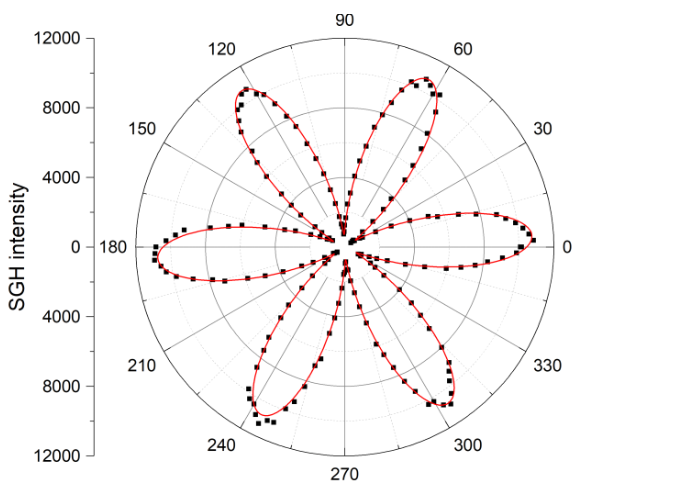}\caption{Polar plot of the polarized SHG signal of CVD grown 3R MoS$_2$ bilayer.}
\label{FigS2}
\end{figure}

Second harmonic generation (SHG) spectroscopy, which is a second-order nonlinear process, has been performed to determine the stacking order of CVD grown MoS$_2$ bilayers. 
Broken inversion symmetry and strong light matter interaction in bilayer 3R MoS$_2$ enables us to observe higher harmonics generation. In a SHG experiment, two photons of the same frequency are converted into a single photon with twice that frequency:
\begin{equation}
\mathbf{P}^{(2)}_i(2\omega) \propto\mathbf{\chi}^{(2)}_{ijk}\mathbf{E}_i(\omega)\mathbf{E}_k(\omega)
\label{eq1}
\end{equation}
where, $\mathbf{P}_i$ is the induced polarization and $\mathbf{E}_{i,k}$ the applied electric field vector components. TMD monolayer crystal lattices in the 2H phase belong to the trigonal prismatic ($D_{3h}$) point group. The second-order nonlinear susceptibility tensor $\chi^{(2)}_{ijk}$ has four non-zero elements with one free parameter $\chi^{(2)}_{0}$ in this symmetry \cite{mennel2019second,leisgang2018optical}. The TMD layers are excited with linearly polarized light under normal incidence. We detect the SHG response with a linear polarizer, oriented along the fundamental polarization angle. This results in a six-fold SHG intensity pattern $I^\parallel_\text{SHG} \propto\cos^23(\theta-\theta_0)$ (SHG polarization parallel to the excitation polarization), where the angle $\theta_0$ is the rotation of the armchair direction of the crystal relative to the laboratory axis. Polarization-resolved SHG experiments were performed using a home-built, confocal microscope set-up at room temperature. A Ti:Sapphire laser source with 76 MHz repetition rate at a wavelength of 804 nm was used. The laser (average power $8$~ mW) with a spot size ($\sim 1.5 ~\mu$m ) was focused on the sample by a microscope objective lens ($NA = 0.65$) at normal incidence and with a fixed linear polarization. The SHG signal was collected by the same objective and directed through a dichroic beamsplitter to a spectrometer with a 300 grooves/mm grating and a nitrogen cooled silicon charge-coupled device (CCD). The SHG intensity strongly depends on the polarization angle ($\theta$ - $\theta_0$) between the laser polarization ${E}(\omega)$ and the armchair direction of the crystal shown in Fig.~\ref{FigS1}. To perform polarization-resolved SHG, the laser polarization was rotated using a half-wave plate, and the SHG intensity $I^\parallel_\text{SHG}$ was collected. The polar plot of 3R MoS$_2$ bilayer in Fig.~\ref{FigS2} directly reveals the stacking order and the crystallographic orientation of the bilayer. It is important to note that for the 2H MoS$_2$ bilayers no SHG signal at all could be detected as inversion symmetry is restored. SHG spectroscopy therefore allows a clear distinction between 3R and 2H stacked bilayers studied in this work.

\subsection*{Optical quality of CVD MoS$_2$ monolayers}

\begin{figure}
\centering
\includegraphics[width=90mm]{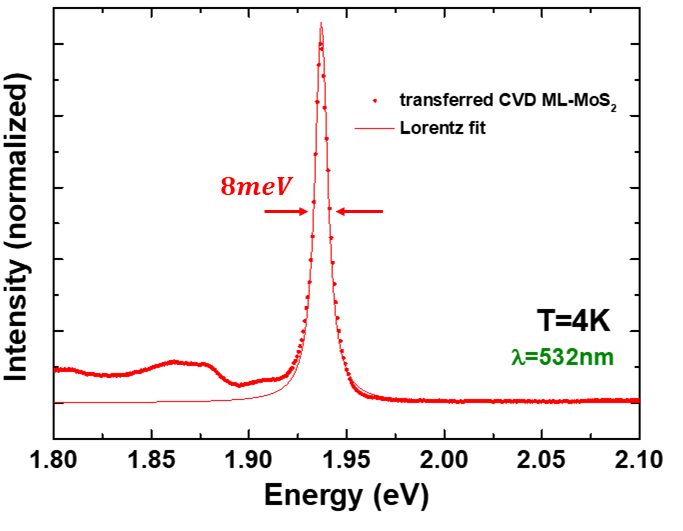}\caption{PL spectrum of ML MoS$_2$ transferred and encapsulated in hBN following the water-assisted transfer method. A laser (532~nm) was used as an excitation source and the spectrum was collected at $T = 4~$K.}
\label{FigS3}
\end{figure}

Photoluminescence (PL) spectroscopy has been employed to evaluate the optical quality of the individual MoS$_2$ MLs after the water-assisted transfer. The PL experiments were carried out in a confocal microscope built in a vibration free, closed cycle cryostat from Attocube at $T= 4K$. The excitation/detection spot diameter of the $532~$nm laser is below $1\mu$m. The optical signal is dispersed in a spectrometer and detected with a Si-CCD camera. In Fig.~\ref{FigS3} we present a typical PL spectrum collected from an hBN-encapsulated ML MoS$_2$ next to the manually assembled bilayers. The A-exciton emission has a linewidth of $8~$meV at the full width at half maximum (FWHM), very close to the recently reported high optical quality encapsulated CVD ML MoS$_2$ \cite{Shree:2019hom}.

\begin{figure*}
\includegraphics[width=0.85\linewidth]{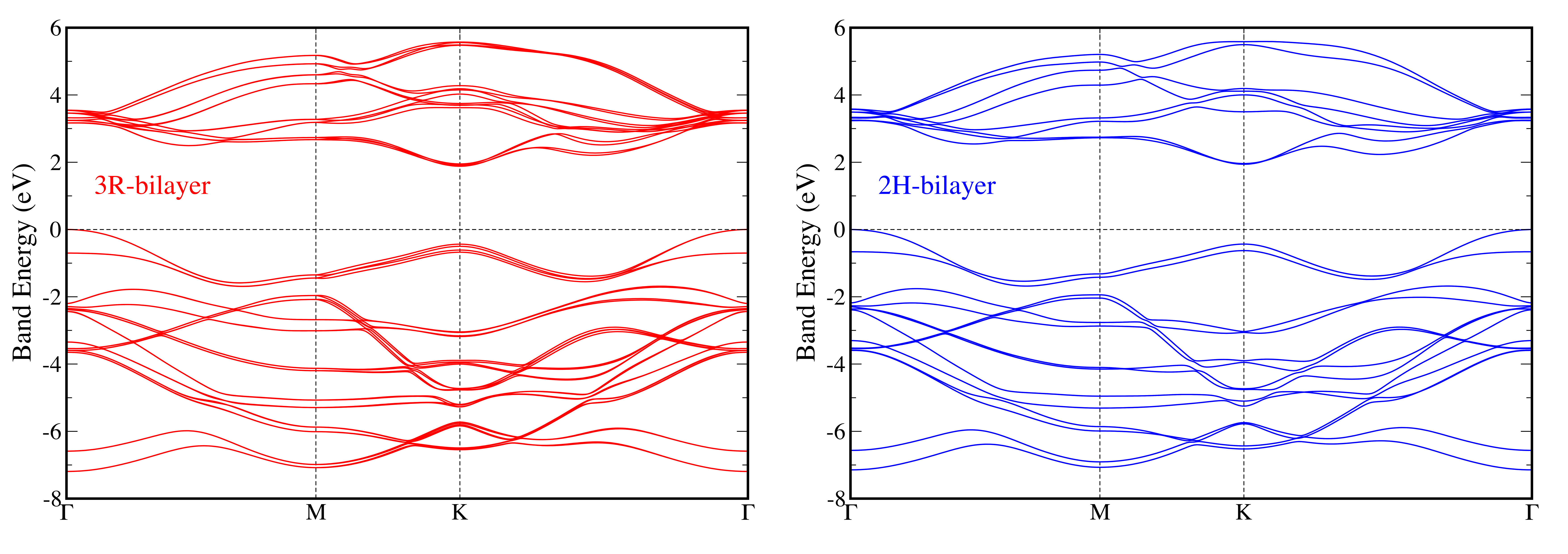}
\caption{\label{FigS4} $G_0W_0$ band structures of 2H and 3R bilayers.}
\end{figure*}

Clear observation of interlayer excitons for CVD grown samples is due to the improved optical quality as compared to earlier studies and due to performing the measurements at cryogenic temperatures. Temperature (i.e. thermal effects), contamination of sample-substrate interface, interference effects and intrinsic sample quality are factors that may have a detrimental contribution on the detection of interlayer excitons in differential reflectivity experiments. The spectral linewidth and the exciton oscillator strength are highly sensitive to the aforementioned factors. Encapsulation of as-grown and manually-assembled CVD bilayers combined with a careful selection of the top and bottom hBN thickness significantly reduces the inhomogeneous broadening and maximizes the oscillator strength of interlayer excitons. This results in high optical quality samples for a precise determination of the interlayer interaction in different stacking orders. \\

\subsection*{DFT-$GW$-BSE calculations}
In the main text we only discuss the conduction and valence states around the K-point of the Brillouin zone, here we show in  Fig.~\ref{FigS4} the band structure calculations over the full Brillouin zone using $\Gamma$-M-K-$\Gamma$ path of 2H and 3R bilayers at the $G_0W_0$ level of theory. The degeneracy leave that appears when going from the 2H to the 3R case is present almost over the entire path.

 \textbf{Acknowledgements}\\ (*) I.P. and S.S. contributed equally to this work. Toulouse acknowledges funding from ANR 2D-vdW-Spin, ANR VallEx, ANR MagicValley, ITN 4PHOTON Marie Sklodowska Curie Grant Agreement No. 721394 and the Institut Universitaire de France. Growth of hexagonal boron nitride crystals was supported by the Elemental Strategy Initiative conducted by MEXT, Japan, and CREST (JPMJCR15F3), JST. I.C.G. thanks the CALMIP initiative for the generous allocation of computational times, through the project p0812, as well as the GENCI-CINES and GENCI-IDRIS for the grant A006096649. For FSU Jena this project received funding from the joint European Union’s Horizon 2020 and DFG research and innovation programme FLAG-ERA under a grant TU149/9-1, DFG Collaborative Research Center SFB 1375 ‘NOA’ Project B2 and the Th\"uringer MWWDG via FGR 0088 2D-Sens. N.L. and R.J.W. acknowledge funding from the PhD School \textit{Quantum Computing and Quantum Technology}, SNF (Project No.$200020_156637$) and NCCR QSIT.\\
 

\end{document}